\documentclass[aps,prl,showpacs,twocolumn]{revtex4}
\usepackage{bbm}
\usepackage{mathrsfs}
\usepackage{epsfig,psfrag}
\usepackage{amsmath,amsfonts,amssymb}
\usepackage[usenames]{color}

\begin{document}

\preprint{draft}

\title{Casimir interaction between a plate and a cylinder}

\author{T. Emig,$^{\rm a}$ R.~L.~Jaffe,$^{\rm b,c}$  M.~Kardar,$^{\rm c}$ and A.~Scardicchio$^{\rm b,c}$\\{~}\\
$^{\rm a}$Institut f\"ur Theoretische Physik, Universit\"at zu
K\"oln, Z\"ulpicher Stra\ss e 77, 50937 K\"oln, Germany\\[1ex]
$^{\rm b}$Center for Theoretical Physics and Laboratory for Nuclear Science\\
$^{\rm c}$Department of Physics\\ Massachusetts Institute of
Technology, Cambridge, MA 02139, USA }

\date{\today}

\begin{abstract}
  We find the exact Casimir force between a plate and a cylinder, a
  geometry intermediate between parallel plates, where the force is
  known exactly, and the plate--sphere, where it is known at large
  separations. The force has an unexpectedly weak decay $\sim L/(H^3
  \ln(H/R))$ at large plate--cylinder separations $H$ ($L$ and $R$ are
  the cylinder length and radius), due to transverse magnetic modes.
  Path integral quantization with a partial wave expansion
  additionally gives a qualitative difference for the density of
  states of electric and magnetic modes, and corrections at finite
  temperatures.
\end{abstract}


\maketitle

With recent advances in the fabrication of electronic and
mechanical systems on the nanometer scale quantum effects like
Casimir forces have become increasingly
important\cite{Cleland+96,CAKBC2001}. These systems can probe
mechanical oscillation modes of quasi one-dimensional structures
such as nano wires or carbon nanotubes with high precision
\cite{Sazonova+04}. However, thorough theoretical investigations
of Casimir forces are to date limited to ``closed'' geometries
such as parallel plates \cite{Casimir48} or, recently, a
rectilinear ``piston'' \cite{hjks}, where the zero point
fluctuations are not diffracted into regions which are
inaccessible to classical rays. A notable exception is the original
work by Casimir and Polder on the interaction between a plate and
an atom (sphere) at asymptotically large separation
\cite{Casimir+48}.

\begin{figure}[ht]
\includegraphics[width=0.9\linewidth]{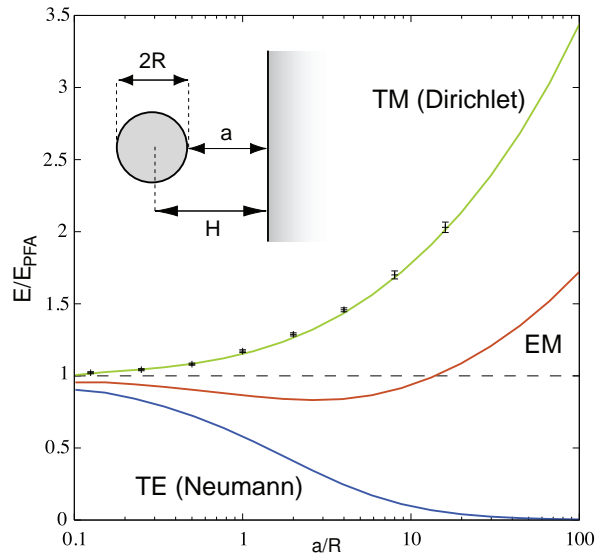}
\vspace*{-.2cm }
\caption{\label{fig:energy}Ratio of the exact
Casimir energy to the PFA
  for the geometry shown in the inset.   All curves are obtained at order
  $25$ of the partial wave expansion, and  the
  accuracy lies within the line thickness even at small $a/R$.
   The Dirichlet data points  are from
  Ref.~\cite{Gies-priv}.}
\end{figure}

In this Letter we consider the electrodynamic Casimir interaction
between a plate and a parallel cylinder (or ``wire''), both
assumed to be perfect metals (see inset of Fig.~\ref{fig:energy}).
We show that the Casimir interaction can be computed without
approximation for this geometry. We believe that the methods
presented here may yield exact solutions for other interesting
geometries as well.  This geometry is also of recent experimental
interest: Keeping two plates parallel has proved very difficult.
The sphere and plate configuration avoids this problem, but the
force is not extensive. The cylinder is easier to hold parallel
and the force is extensive in its length \cite{Brown+05}.

Casimir interactions, while attractive for perfect metals in all known
cases, depend strongly on geometry.  Consider the Casimir interaction
energy (discarding separation independent terms) at asymptotically
large $H$ for three fundamental geometries which differ in the
co-dimension of the surfaces\cite{Scardicchio05}: two plates,
plate--cylinder, and finally, plate--sphere, corresponding to
co-dimension 1, 2, and 3, respectively.  It is instructive to consider
both a scalar field which vanishes on the surfaces (D $\equiv$
Dirichlet) and the electromagnetic field (EM). For parallel plates
(area $A$) $E\sim \hbar c A/H^3$ in both cases \cite{Casimir48}.  For
a plate and a sphere of radius $R$, $E\sim \hbar c R/H^2$
\cite{Bulgac+05} for the Dirichlet case, as compared with $E\sim \hbar
c R^3/H^4$ for the EM case \cite{Casimir+48}. Based on these results,
expectations for the plate and cylinder geometry might range from
$\sim \hbar c LR^2/H^4$, proportional to the cylinder volume, to $\sim
\hbar c LR/H^3$, proportional to its surface area, or even $\sim \hbar
c L/H^2$ with a potential non-power law dependence on the radius.

A simple but uncontrolled method for study of non-planar geometries is
the proximity force approximation (PFA), where the system is treated
as a sum of infinitesmal parallel plates \cite{Bordag+01}.  Applied to
the plate--cylinder geometry, the PFA yields $E_{\rm PFA} =
-\frac{1}{960} \pi^{3}\hbar c L \sqrt{R/2a^{5}}$ to leading order in
$a/R$, where $a=H-R$. Other approximations include semi-classical
methods based on the Gutzwiller trace formula \cite{Schaden+00}, and a
recent optical approach which sums also over closed but
  non-periodic paths\cite{Jaffe+04}.  For large separations, a
multiple scattering approach is available \cite{Balian}, but has not
been adapted to this geometry.  For the Dirichlet case, a Monte Carlo
approach based on worldline techniques has been applied to the
plate-cylinder case \cite{Gies+03}.

Our result provides a test for the validity of these approximate
schemes, and also provides insight into the large distance limit.  In
particular, we find the unexpected result that the electrodynamic
Casimir force for the plate--cylinder geometry has the weakest of the
possible decays,
\begin{equation}
\label{eq:force-result}
F = - \frac{1}{8\pi} \frac{\hbar cL}{H^3 \ln(H/R)} \, ,
\end{equation}
as $H/R\to\infty$.  The same asymptotic result applies to a scalar
field with Dirichlet boundary conditions.  
Interestingly,  the decay exponent of the force
\emph{is not monotonic} in the number of co-dimensions: ($-4$,
$-3-\epsilon$, $-5$) for co-dimension (1,2,3) respectively.
In contrast the Dirichlet case is monotonic with exponents
($-4$,$-3-\epsilon$, $-3$).

In the remainder we derive these results, summarized in
Eqs.~(\ref{eq:E-general})-(\ref{eq:N-matrix}), using path integral
techniques. Our approach also yields the distance dependent part
of the density of states, which contains the complete geometry
dependent information of the photon spectrum, and is useful
for computing thermal contributions to the force.

The translational symmetry along the cylinder axis enables a
decomposition of the EM field into transverse magnetic (TM) and
electric (TE) modes \cite{Emig+01} which are described by a scalar
field obeying Dirichlet (D) or Neumann (N) boundary conditions
respectively.  We can compare our TM results to recent Monte Carlo
Dirichlet results \cite{Gies+03}.  Moreover, the mode
decomposition turns out to be useful also in identifying the
physical mechanism behind the weak decay of the force, which at
large distance is fully dominated by D modes.

Our starting point is a path integral representation \cite{LK91} for the effective action 
which yields a trace formula for the density of states (DOS) \cite{Buscher+05}. 
The latter is then evaluated using a partial wave expansion. 
The DOS on the imaginary frequency axis  is related to a Green's function
by $\rho(iq_0)=(2q_0/\pi)\int d^3 {\bf x} \, G({\bf x},{\bf
  x};q_0)$, where $G({\bf x},{\bf x}';q_0)$ is the Green's
  function for the scalar field with action $S=\frac{1}{2}\int d^3x
  (|\nabla\Phi|^2 +(q^{0})^{2}\Phi^{2})$. The effect of boundaries on
the Green function  can be obtained by placing functional delta
functions on the boundary surfaces in the functional
integral\cite{LK91}. By integrating out both the field $\Phi$ and
the auxiliary fields which represent the delta functions on the
surfaces, one obtains the trace formula \cite{Buscher+05}
\begin{equation}
\label{eq:trace-formula}
\delta\rho(q_0)=-\frac{1}{\pi} \frac{\partial}{\partial q_0} {\rm
Tr} \ln \left( M M_\infty^{-1}\right) \, ,
\end{equation}
where  $\delta\rho(q_0)$ is the change in the DOS caused by moving
the plate and cylinder in from infinity.  The information about
geometry is contained in the matrix $M$ of the quadratic
action for the auxiliary fields, given by
$M_{\alpha\beta}({\bf u},{\bf u}';q_0)=G_0({\bf s}_\alpha({\bf
  u})-{\bf s}_\beta({\bf u}');q_0)$ for D and by
$M_{\alpha\beta}({\bf u},{\bf u}';q_0)=\partial_{{\bf
n}_\alpha({\bf
    u})}\partial_{{\bf n}_\beta({\bf u}')}G_0({\bf s}_\alpha({\bf
  u})-{\bf s}_\beta({\bf u}');q_0)$ for N boundary conditions;
$G_0=e^{-q_0|{\bf x}|}/ 4\pi|{\bf x}| $ is the free space Green
function, $\partial_{{\bf n}_\alpha}$ is its derivative normal to
the surface, and ${\bf s}_\alpha({\bf u})$ parametrizes the
surfaces (which are numbered by $\alpha=1$, $2$) in terms of
surface coordinates ${\bf u}$.  $M_\infty^{-1}$ is the functional
inverse of $M$ at infinite surface separation.  The trace in
Eq.~\eqref{eq:trace-formula} runs over ${\bf u}$ and $\alpha$. For
the cylinder with its axis oriented along the $x_1$ direction we
set ${\bf s}_1(x_1,\varphi)=(x_1,R\sin(\varphi),R\cos(\varphi))$
and for the plate ${\bf s}_2(x_1,x_2)=(x_1,x_2,H)$ (see inset of
Fig.~\ref{fig:energy}).

The Casimir energy of interaction is given by $E=(\hbar c/2)
\int_0^\infty dq_0 q_0 \delta\rho(q_0)$. After transforming to
momentum space, $\tilde M$, the Fourier transform of the matrix
$M$ has block diagonal form with respect to $q_0$ and the momentum
$q_1$ along the cylinder axis, so the Casimir energy can be
expressed as,
\begin{equation}
E =\frac{\hbar cL}{8 \pi^2} \int\!\!\!\int dq_0 dq_1 \ln
\frac{\det \tilde M(q_0,q_1)}{\det \tilde M_\infty(q_0,q_1)} \, .
\end{equation}
The elements of the matrix $M$ are labeled by the integer
index $m=-\infty,\ldots,\infty$ coming from the compact
$\varphi$-dimension of the cylinder, and the momentum $q_2$ along
the other direction parallel to the plate, to read
\begin{equation}
\tilde M = \left(
\begin{array}{cc}
A_{[m,m'] }& B_{[m,q_2']} \\
B^T_{[q_2,m']} & C_{[q_2,q_2']}
\end{array}
\right).
\end{equation}
The matrix $A_{[m,m']}$ is diagonal,  with elements $A_{[m,m]}\equiv A_m=I_m(ru)K_m(ru)$
for D and $A_m=(u/H)^2I'_m(ru)K'_m(ru)$ for N modes.
The matrix $C$ also has only diagonal
elements $C_{[q_2,q_2]}\equiv C(q_2) = H/(2\sqrt{u^2+u_2^2})$ for D and
$C(q_2) = -\sqrt{u^2+u_2^2}/(2H)$ for N modes. 
The off-diagonal matrix $B$ is non-diagonal with
$B_{[m,q_2]}\equiv B_m(q_2)=\pi H
e^{-\sqrt{u^2+u_2^2}}I_m(ru)[u/(u_2+\sqrt{u^2+u_2^2})]^m/\sqrt{u^2+u_2^2}
$ for D and $B_m(q_2)=(\pi u/H)
e^{-\sqrt{u^2+u_2^2}}I'_m(ru)[u/(u_2+\sqrt{u^2+u_2^2})]^{-m}$ for N modes.  
Here, we have defined the dimensionless combinations 
$u=H\sqrt{q_0^2+q_1^2}$, $u_2=Hq_2$ and $r=R/H$.
The determinant can be obtained straightforwardly, and the total energy can be 
decomposed to the sum of D and N mode contributions, as
\begin{equation}
\label{eq:E-general}
E  = -\frac{\hbar cL}{H^2} \left[\Phi^D (r) +\Phi^N(r) \right],
\end{equation}
with
\begin{equation}
\label{eq:phi-formula}
\Phi^X(r)=-\frac{1}{4\pi} \int_0^\infty du \, u \ln \left[ \det
(\mathbbm{1} -N^X(u,r))\right] \, .
\end{equation}
The matrix $N^X(u)$ is given in terms of Bessel functions,
\begin{equation}
\label{eq:D-matrix}
N^D_{\mu\nu}(u,r)=\frac{I_\nu(ru)}{K_\mu(ru)} K_{\mu+\nu}(2u),
\end{equation}
for D modes and
\begin{equation}
\label{eq:N-matrix}
N^N_{\mu\nu}(u,r)=-\frac{I'_\nu(ru)}{K'_\mu(ru)} K_{\mu+\nu}(2u),
\end{equation}
for N modes. The determinant in Eq.~\eqref{eq:phi-formula} is
taken with respect to the integer indices $\mu$,
$\nu=-\infty,\ldots,\infty$.  If the matrix $N^X$ is restricted to
dimension $(2l+1)$ with $N_{00}^X$ as the central element, it then
describes the contribution from $l$ partial waves, beginning with
$s$-waves for $l=0$.

From  Eq.~\eqref{eq:phi-formula}, one can easily
extract the asymptotic large distance behavior of the energy for
$r=R/H \ll 1$. For Dirichlet modes s-waves dominate, while for
Neumann modes both $s$- and $p$-waves ($l=1$) contribute at
leading order in $r$. The two cases differ qualitatively, with
\begin{equation}
\label{eq:asympt}
\Phi^D(r)=-\frac{1}{16\pi} \frac{1}{\ln r}, \quad{\rm and}\quad
\Phi^N(r)=\frac{5}{32\pi} r^2.
\end{equation}
For $H\gg R$ the EM Casimir interaction is
dominated by the D (TM) modes. Note that a
naive application of the PFA for small $r$, where it is not
justified, yields the incorrect scalings
$\Phi^D(r)=\Phi^N(r)\sim r$.

The natural expectation from the Casimir--Polder result for the
plate-sphere interaction, that the force at large distance is
proportional to the volume of the cylinder, is incorrect.  The
physical reason for this difference is explained by considering
spontaneous charge fluctuations.  On a sphere, the positive and
negative charges can be separated by at most distances of order $R\ll
H$.  The retarded van der Waals interactions between these dipoles and
their images on the plate leads to the Casimir--Polder interaction
\cite{Casimir+48}.  In the cylinder, fluctuations of charge along the
axis of the cylinder can create arbitrary large positively (or
negatively) charged regions.  The retarded interaction of these
charges (not dipoles) with their images gives the dominant term of the
Casimir force.  This interpretation is consistent with the difference
between the two types of modes, since for N modes such charge
modulations cannot occur due to the absence of an electric field along
the cylinder axis.  Eventually, for a finite cylinder, in the very far
region $H\gg L$, the charge fluctuations can be considered again as
small dipoles, and the Casimir--Polder law is expected to reappear,
making the force proportional to the volume of the cylinder $LR^2$.

We next consider arbitrary separations, and use
Eq.~\eqref{eq:phi-formula} to obtain the contribution from higher
order partial waves. A numerical evaluation of the determinant is
straightforward, and we find that down to even small separations
of $a/R=0.1$ the energy converges at order $l=25$, whereas for
$a/R \gtrsim 1$ convergence is achieved for $l=4$.
Fig.~\ref{fig:energy} shows our results for Dirichlet and Neumann
modes and for their
  sum which is  the EM Casimir energy, all
scaled by the corresponding $E_{\rm PFA}$ given above
\cite{foot1}. Both types of modes show a strong deviation from the
PFA for $a/R \gtrsim 1$, especially the Dirichlet energy.
Fig.~\ref{fig:energy} shows also very recent wordline-based Monte
Carlo results for the Dirichlet case at moderate separations
\cite{Gies-priv}, which agree nicely with our exact results.

Eq.~\eqref{eq:asympt} indicates that the Dirichlet dominated
force vanishes logarithmically as $R \to 0$ at fixed $H$.  A
similar result is obtained when the cylinder is replaced by an
infinitesimal thin wire, but an UV cutoff is introduced to control
short wavelength modes.  Both results are a consequence of the fact
that the asymptotic form of Eq.~\eqref{eq:asympt} is {\it independent}
of the actual shape of the cross section of the wire, and the cutoff
$R$ can be identified with any typical scale of the cross section.
The leading asymptotic term in Eq.~\eqref{eq:asympt} is also obtained
\cite{Scardicchio05} from the $s$-wave scattering amplitude for the
2-dimensional problem of a strongly repulsive potential concentrated
on the wire.

\begin{figure}[ht]
\includegraphics[width=.77\linewidth]{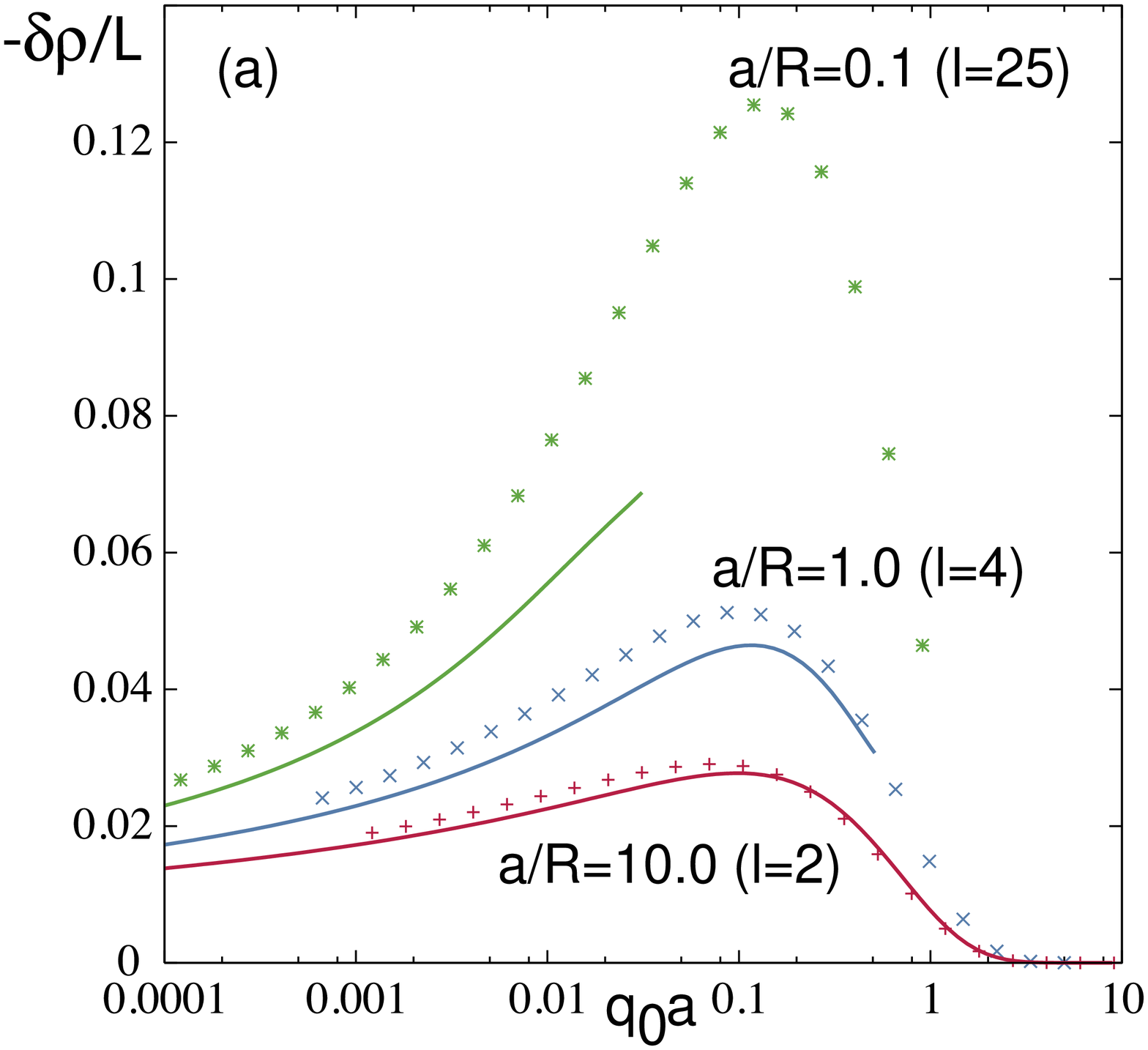}
\includegraphics[width=.77\linewidth]{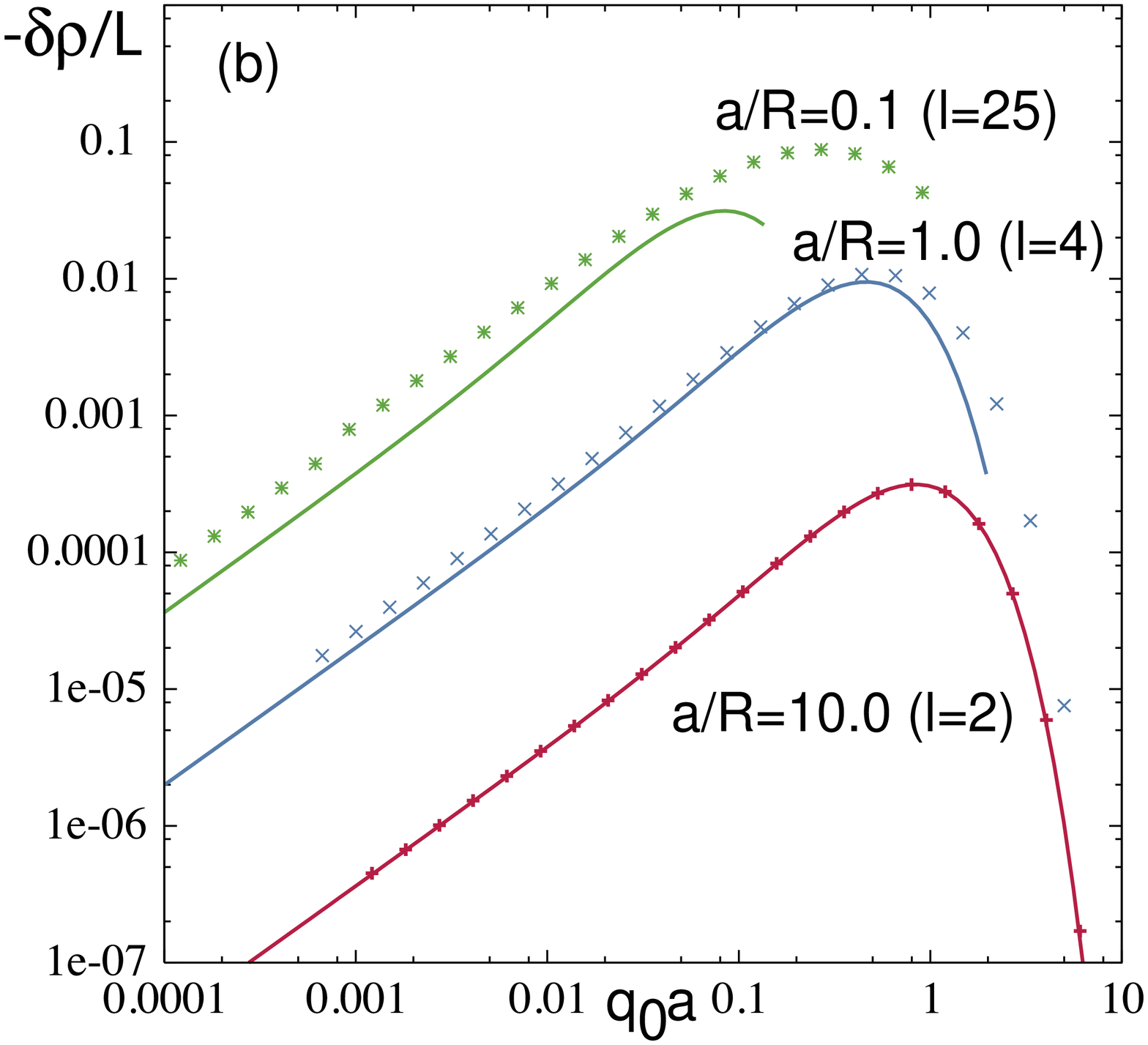}
\vspace*{-0.4cm}
\caption{\label{fig:dos} The change in the density of states for (a)
  Dirichlet and (b) Neumann modes obtained from
  Eq.~\eqref{eq:dos-general} at order $l$.  The solid curves show the
  small $R$ expansion of Eq.~(\ref{eq:dos_asympt}).}
\vspace*{-.7cm} 
\end{figure}

The difference between the D and N modes also appears in the
density of states, which in turn affects the temperature
dependence of the Casimir force.  From
Eq.~\eqref{eq:trace-formula} we obtain the expression
\begin{equation}
\label{eq:dos-general}
\delta\rho_X\!(q_0) \!= \!-\frac{q_0HL}{\pi^2} \!\!\!\int_0^\infty\!\!
\frac{du}{u^2} \ln\frac{\det\left[\mathbbm{1}
\!\!-\! \! N^X(\sqrt{u^2+(q_0H)^2})\right]} {\det\left[
\mathbbm{1}-N^X(q_0H)\right]} \, ,
\end{equation}
which is convenient both for numerical and analytic computations.
Numerical evaluation yields the results shown in
Fig.~\ref{fig:dos} for general values of $R$.  
Analytical results in the limit of small $R/H$ are obtained by
considering only the $s$-waves  for Dirichlet modes, and the
$s$- and $p$-waves for Neumann modes.  
For Dirichlet modes we expand in $1/\ln(q_0 R)$,
whereas for Neumann modes the small parameter is $r=R/H$. 
To leading order we find
\begin{subequations}
\label{eq:dos_asympt}
\begin{eqnarray}
\!\!\!\!\!\!    \delta\rho_{\rm D}(q_0)  \!\! &=& \!\!
\frac{L}{2\pi} \frac{e^{-2q_0 H}}{\ln(q_0 R)}
 + {\cal O}\left(\ln^{-2}(q_0 R)\right)\, , \\
\!\!\!\!\!\!  \delta\rho_{\rm N}(q_0) \!\! &=& \!\! -\frac{q_0 HL}{8\pi} (1+6
q_0H)e^{-2q_0 H} r^2 + {\cal O}(r^3).
\end{eqnarray}
\end{subequations}
Fig.~\ref{fig:dos} allows for an assessment of the validity range
of the expansions of Eq.~\eqref{eq:dos_asympt} which are shown as
solid curves.

These results for the DOS allow us to evaluate for the first time
finite temperature contributions to the Casimir interaction in an open
geometry. The difference between the free energy ${\cal F}$ and
the Casimir energy at $T=0$ can be written as \cite{Balian} ($k_B$
is Boltzmann constant)
\begin{equation}
\delta {\cal F}={\cal F}-E=\pi\, k_B T \int_0^\infty dq_0 \,
g(q_0) \delta\rho(q_0),
\end{equation}
with the function $g(q_0)=\sum_{k=1}^\infty \sin(2\pi k
q_0\lambda_T)/(\pi k)$ and $\lambda_T=\hbar c/(2\pi k_B T)$. In the
limit $R \ll (H,\lambda_T)$ but for general
  $H/\lambda_T$, we can use the expansion of
Eq.~\eqref{eq:dos_asympt} to obtain to leading order in
  $1/\ln(R/\lambda_T)$ and $R/H$ for D and N modes, respectively, the
thermal contributions
\begin{subequations}
\label{eq:finite-T}
\begin{eqnarray}
 \delta {\cal F}_{\rm D}  &=& \frac{k_BT}{8}\frac{L}{\ln(R/\lambda_T)H} \left[
\coth\left(\frac{H}{\lambda_T}\right)-\frac{\lambda_T}{H}\right] \, ,\\
 \delta {\cal F}_{\rm N} &=& - \frac{k_BT}{64} \frac{L\lambda_TR^2}{H^4}\bigg[
7\frac{H}{\lambda_T} \coth\left(\frac{H}{\lambda_T}\right) \\
 && +\frac{7(H/\lambda_T)^2}{\sinh^2(H/\lambda_T)}+6
\left(\frac{H}{\lambda_T}\right)^3
\frac{\cosh(H/\lambda_T)}{\sinh^3(H/\lambda_T)}-20 \bigg]
\nonumber \, .
\end{eqnarray}
\end{subequations}
It is interesting to note that $\delta {\cal F}_{\rm N}$ has a minimum
at $H/\lambda_T=2.915\ldots$, where the corresponding thermal force
changes from repulsive at small $H$ to attractive at large $H$.  At
low temperatures, the finite $T$ contributions to the Casimir force
$\delta F=-\partial \delta {\cal F}/\partial H$,
\begin{subequations}
\label{eq:low-T}
\begin{eqnarray}
 \delta F_{\rm D}  &=& \frac{2\pi^3}{45}k_BT\left(\frac{k_BT}{\hbar c}\right)^3 \frac{HL}{
\ln(R/\lambda_T)}\, , \\
 \delta F_{\rm N}   &=&
\frac{64 \pi^5}{945}k_BT \left(\frac{k_BT}{\hbar c}\right)^5 R^2 HL\, ,
\end{eqnarray}
\end{subequations}
have to be added to Eq.~\eqref{eq:force-result} for $H \ll \lambda_T$.
At larger temperatures with $R \ll \lambda_T \ll H$, one has the
scalings $\delta F_{\rm D} \sim k_BT LH^{-2}/\ln(R/\lambda_T)$ and
$\delta F_{\rm N} \sim -k_BTLR^2H^{-4}$.  At the extreme high
temperature limit of $\lambda_T\ll R$, only thermal fluctuations
remain, and $\hbar$ should disappear from the equations.  This
`classical limit' is well known for parallel plates \cite{Milonni} and
is obtained for smooth, arbitrary geometries within the multiple
scattering approach \cite{Balian}, and the optical approximation
\cite{Scardicchio:2005di}.  (Note that for the D modes a subleading
$\hbar$ still survives in the logarithm.)

Finally, we note that our approach can be extended also to
multiple wires and distorted beams. Our results should be relevant
to nano-systems composed of 1-dimensional structures and also to
other types of fields as, e.g., thermal order parameter
fluctuations.

We thank H. Gies for discussions and especially for providing the
data of Fig.~\ref{fig:energy} prior to publication.
This work was supported by the DFG
through grant EM70/2 (TE), the Istituto
Nazionale di Fisica Nulceare (AS), the NSF through grant
DMR-04-26677 (MK), and the U.~S.~Department of Energy (D.O.E.)
under cooperative research agreement \#DF-FC02-94ER40818 (RLJ \&
AS).


\begin{thebibliography}{99}
\bibitem{foot1} The energies $E_{\rm PFA}$ for D and N modes
are both half of the EM PFA energy.
\end{thebibliography}
\end{document}